\documentclass[aps,pre,twocolumn,superscriptaddress]{revtex4}

\usepackage{graphicx}
\usepackage{braket}
\usepackage[driverfallback=dvipdfm]{hyperref}
\hypersetup{pdfpagemode=FullScreen,colorlinks=true,breaklinks,urlcolor=blue,linkcolor=blue,citecolor=blue}

\usepackage{ulem,CJK}

\usepackage{color}
\usepackage{cleveref}

\newcommand{\comment}[1]{}

\newcommand{\fract}[1]{\frac{#1}{2}}
\newcommand{\lr}[1]{\left( #1\right)}
\newcommand{\figwid}{8.3 cm}

\begin{document}


\title{Wannier basis method for KAM effect in quantum mechanics}


\begin{CJK}{UTF8}{gbsn}
\author{Chao Yin (尹超)}
\affiliation{International Center for Quantum Materials, School of Physics, Peking University, Beijing 100871, China}
\author{Yu Chen (陈宇)}
\affiliation{Center for Theoretical Physics, Department of Physics, Capital Normal University, Beijing 100048, China}
\author{Biao Wu (吴飙)}
\affiliation{International Center for Quantum Materials, School of Physics, Peking University, Beijing 100871, China}
\affiliation{Collaborative Innovation Center of Quantum Matter, Beijing 100871, China}
\affiliation{Wilczek Quantum Center, School of Physics and Astronomy, Shanghai Jiao Tong University, Shanghai 200240, China}


\date{\today}

\begin{abstract}
The effect of Kolmogorov-Arnold-Moser (KAM) theorem in quantum systems is
manifested in dividing eigenstates  into regular and irregular states. We propose an effective method
based on Wannier basis in phase space to illustrate this division of eigenstates. The quantum kicked-rotor
model is used to illustrate this method, which allows us to define the area and effective dimension 
of each eigenstate 
to distinguish quantitatively regular and irregular eigenstates. This Wannier basis method also allows us to define the length of a Planck cell in the spectrum that measures how many Planck cells the system will traverse if it starts at the given Planck cell.
Moreover, with this Wannier approach, we are able to   clarify the distinction between KAM effect and Anderson localization.
\end{abstract}


\maketitle
\end{CJK}
\section{Introduction}
There are two contrasting types of motion in classical dynamics. The first type  is regular orbits in integrable systems,
where there exist $N$ independent conserved quantities ($N$ is the degree of freedom) that restrict motion 
to an $N$-dimensional torus  in phase space\cite{arnol2013mathematical}. The second type is irregular motion in chaotic systems,
where most orbits explore almost all points in a $2N-1$ dimensional energy surface in the sense of ergodicity and mixing\cite{Gutzwiller}.
According to the well-known Kolmogorov-Arnold-Moser (KAM) theorem
\cite{kolmogorov1954conservation,Moser,arnold1963vi}, there is a smooth crossover from an integrable system to a chaotic
system. Specifically, Kolmogorov, Arnold, and Moser  considered a Hamiltonian of the form $H = H_0 +\epsilon H'$,
where $H_0$ is integrable. They found that  a subset of the torus solutions under $H_0$ are deformed and survive under a sufficiently small perturbation $\epsilon H'$; while motion near the unstable tori is chaotic and fills regions with dimensionality $2N-1$. As a result, the phase space is divided into integrable and chaotic regions, with the measure of the latter
growing with $\epsilon$.

As classical dynamics is the semi-classical limit ($\hbar\rightarrow 0$) of quantum dynamics,
one expects similar KAM effects in quantum mechanics. There has been lots of work extending KAM to quantum systems \cite{KAM1983PhysRevLett.51.947, Hose_1984, KAMbarrier_PhysRevLett.57.2883, Reichl1987, PhysRevA.37.3972, Evans2004, Grebert2011, Pol_RevModPhys.83.863, glimmer_PhysRevX.5.041043}, especially, KAM in quantum many-body systems has become a recent interest \cite{Pol_RevModPhys.83.863, glimmer_PhysRevX.5.041043}. In this paper we focus on cases 
that have  classical limits. In these systems,  previous studies have shown that quantum
KAM effects are manifested in eigen-energies and eigenfunctions. For systems of KAM types,
both their eigen-energies and eigenfunctions have two parts: regular part and irregular part \cite{0022-3700-6-9-002,AIHPA_1976__24_1_31_0,berry1977,berry1983semiclassical,voros1979stochastic}.
In particular,  to quantitatively understand regular and irregular eigenfunctions,
there have been serious efforts to  compare quantum eigenfunctions to classical orbits in phase space
either using Wigner distribution \cite{berry1977,berry1983semiclassical} or  Husimi distribution \cite{scar_PhysRevLett.53.1515,Husimi_PhysRevA.35.3546,newclass_PhysRevLett.85.1214}.

In this work we propose a different method to capture the quantum KAM effect, i.e., the division of regular and irregular eigenstates.
In our approach, we divide the  phase space into Planck cells and assign a Wannier function to each Planck cell~\cite{HanPRE,Wannier1742-5468-2018-2-023113,Jiang}. These Wannier functions
form an orthonormal and complete basis and they allow us to project a wave function unitarily to phase space .
With this unitary projection, we are able to define for every eigenfunction an area, which measures
how much the eigenfunction occupies in the phase space. We are also able to define an
effective dimension for every eigenfunction. Our numerical results show that the effective dimension
of an irregular eigenfunction is the same as the phase space while a regular eigenfunction has a lower dimension. We are also able to define a length for each Planck cell  by projecting Wannier basis back to the eigenstates. We argue with numerical evidence that this length measures how much phase space the long time quantum trajectory will traverse when starting from the given Planck cell.

We illustrate our method using the quantum kicked-rotor (QKR) model, whose classical counterpart,
the classical kicked-rotor (CKR) \cite{CHIRIKOV1979263}, is one of the simplest models governed by the KAM theorem.
We first consider the case of  $\hbar_e/2\pi$ being a rational number , where $\hbar_e$ is the effective Planck constant. 
Then we extend our discussion to generic $\hbar_e$ and show the distinction between KAM effects and Anderson localization.
\section{QKR Model and the Wannier basis approach}
\subsection{QKR model}
The dimensionless Hamiltonian of the QKR can be written as \cite{Jiang}
\begin{equation}\label{hamil}
\hat{H} = \fract{\hat{p}^2} + K \cos{\hat{x}} \sum_{j=-\infty}^{+\infty} \delta(t-j),
\end{equation}
where $\hat{p}$ is the dimensionless angular-momentum operator, $\hat{x}$ is the angular
coordinate operator, $t$ is the dimensionless time, and $K$ is the kicking strength. In the coordinate representation, $\hat{p}=-i\hbar_e(\partial / \partial x)$, where $\hbar_e$ is the dimensionless effective Planck constant. The dimensionless Schr\"{o}dinger
equation is $i\hbar_e(\partial / \partial t)|\Psi\rangle=\hat{H}|\Psi\rangle$.  Note that for a real rotor with moment of inertia $I$ and driving period $T$, the effective Planck constant is given by $\hbar_e=\hbar T/I$.

The evolution operator over one period is
\begin{equation}
\hat{U} = \exp\lr{-\frac{i}{2}\frac{\hat{p}^2}{\hbar_e} } \exp\lr{-\frac{i}{\hbar_e}K\cos \hat{x} }\,.
\end{equation}
For this system, the momentum basis $\braket{x|n}=e^{inx}$ ($n$ is an integer) is the most convenient. The
matrix elements of  $\hat{U}$ are given by
\begin{equation}
U_{n'n}\equiv\langle n'| \hat{U} |n\rangle = (-i)^{n-n'}J_{n-n'}(\frac{K}{\hbar_e}) \exp\lr{-\frac{i n'^2 \hbar_e}{2} },
\end{equation}
where $J_{n-n'}(K/\hbar_e)$ is the first kind Bessel function. 
The eigenstates of the Floquet operator $\hat{U}$ in this periodically-driven system 
play the same role as energy eigenstates in a time-independent system.

In the following discussion, unless specified otherwise, we focus on the case that 
$\hbar_e/(2\pi)$ is rational, that is, $\hbar_e=2\pi M/N$, where $M,N$ are coprime positive integers\cite{PRAChang}. 
This is called quantum resonance \cite{qr_casati}. In this work for simplicity we assume that $N$ is even.  
For even $N$,  we find that  $U_{n+N\ell, n'+N\ell}=U_{n n'}, (l=0,\pm 1,\cdots)$, which reflects 
a translational symmetry in $p$ space (see Appendix~\ref{reson} for details). This means that an eigenstate  $|\phi\rangle$ of 
the unitary operator $\hat{U}$ must be of the form of Bloch states 
\begin{equation}\label{eq:bloch}
\phi(s+N\ell) \equiv\langle s+Nl |\phi\rangle = e^{-il\theta}\phi_\theta(s),\qquad 0\le \theta < 2\pi,
\end{equation}
where $s=1,\cdots,N,\ l=0,\pm 1,\cdots$, and $\theta$ is a Bloch wave vector along $p$. This shows that all eigenstates are extended in $p$ space, and thus have an infinite expectation value of kinetic energy. Moreover, $\phi_\theta(s)$ is the eigenstate of a $N\times N$ matrix $V_\theta$:
\begin{eqnarray}\label{eq:eigen}
  \sum_{s'=1}^{N} V_\theta(s,s') \phi_\theta(s') &=& e^{-i\omega_\phi}\phi_\theta(s), 
\end{eqnarray}
where $\omega_\phi$ is the quasi-energy of $|\phi\rangle$, and
\begin{eqnarray}\label{eq:V}
V_\theta(s,s') &\equiv& \sum_{\ell'=-\infty}^{+\infty} U_{s,s'+N\ell'} e^{-i\ell'\theta}.
\end{eqnarray}

This suggests that the Hilbert space can be reduced naturally to finite-dimensions without  truncation, which is one of the benefits of the resonance condition. Our results have little dependence on the Bloch wave vector $\theta$, which is also shown in \cite{PRAChang}. Therefore,  
we will always choose $\theta=0$, and denote $V_\theta$ simply by $V$ and $\phi_\theta$ by $\phi$. The second benefit with the resonance condition 
is that the quantum phase space is naturally constructed, while there is some insignificant ambiguity 
when $\hbar_e$ is generic, which we will see in the next section.

\subsection{Construction of quantum phase space}
In order to compare quantum dynamics with its classical counterpart, we construct a quantum phase space. 
This is accomplished by dividing the classical phase space into Planck cells and assigning a Wannier function 
to each Planck cell~\cite{HanPRE,Wannier1742-5468-2018-2-023113,Jiang}. 
These Wannier functions  are localized in their corresponding Planck cells  and form a complete basis for the Hilbert space. 
In this work, we follow the method in Ref.\cite{Jiang}.
Suppose $N=N_x\cdot N_p$, where $N_x,N_p$ are integers. The Wannier function is constructed as follows 
\begin{equation}\label{eqn:XP}
|\mathcal{X},\mathcal{P}\rangle = \frac{1}{\sqrt{N_x}} \sum_{n=1}^{N_x} \exp\lr{-i\frac{2\pi \mathcal{X} n}{N_x}} |n+\mathcal{P}N_x\rangle,
\end{equation}
where $\mathcal{X}=0,1,\cdots,N_x-1$ and $\mathcal{P}=0,1,\cdots,N_p-1$. It is straightforward to show that the new basis are orthonormal and complete. From Eq.~(\ref{eqn:XP}), it is clear that $|\mathcal{X},\mathcal{P}\rangle$ is localized in $p$ space. Moreover, it is also localized in $x$ space because its $x$ representation is given by
\begin{eqnarray}
\langle x|\mathcal{X},\mathcal{P}\rangle &=& \frac{1}{\sqrt{2\pi N_x}} \frac{\sin(N_xx/2)}{\sin\lr{\frac{x}{2}-\pi\frac{\mathcal{X}}{N_x}}} \nonumber \\ &\ &\exp\left[ \frac{i}{2} (2\mathcal{P}N_x-N_x+1)x-i\pi\frac{\mathcal{X}}{N_x} \right], 
\end{eqnarray}
whose norm is plotted in \cite{Jiang}.

Thus  any quantum state $|\psi\rangle$ has a phase space representation 
$|\psi\rangle = \sum|\mathcal{X},\mathcal{P}\rangle\langle\mathcal{X},\mathcal{P}|\psi\rangle $, and $P_{\mathcal{X},\mathcal{P}}=|\langle\mathcal{X},\mathcal{P}|\psi\rangle|^2$ is the
probability for $|\psi\rangle$ to be in Planck cell $(\mathcal{X},\mathcal{P})$. We emphasize that this basis can be constructed as long as  one has the classical action-angle pairs $(p,x)$, where $x$ has periodic boundary condition. If the natural coordinate of the classical system is not the angle variable, one can also numerically obtain an orthonormal and complete basis of Wannier functions efficiently \cite{Wannier1742-5468-2018-2-023113}.

If we push the limit $N_x, N_p \rightarrow \infty$ keeping $M$ constant, we get an unlimited resolution in the phase space: $\hbar_e\rightarrow 0$, and it can be proved that the quantum dynamics will be reduced to the standard map for the CKR\cite{Jiang}, that is, $\langle \mathcal{X},\mathcal{P}|\hat{V}| \mathcal{X}_0,\mathcal{P}_0 \rangle$ will vanish unless
\begin{eqnarray}\label{eq:stan_map}
\bar{\mathcal{P}} &=& \bar{\mathcal{P}_0} + \frac{K}{2\pi M} \sin\lr{2\pi \bar{\mathcal{X}_0} }, \\
\bar{\mathcal{X}} &=& \bar{\mathcal{X}_0} + M \bar{\mathcal{P}},
\end{eqnarray}
where $\bar{\mathcal{X}}=\mathcal{X}/N_x \in [0,1], \bar{\mathcal{P}}=\mathcal{P}/N_p \in [0,1]$. Taking $\bar{\mathcal{P}}' = M \bar{\mathcal{P}}$, one can see that the map for the pair $(\bar{\mathcal{P}}', \bar{\mathcal{X}})$ is exactly the standard map in CKR. The effect of $M$ is to divide the phase space $0<\bar{\mathcal{X}}<1, 0<\bar{\mathcal{P}}<1$ into $M$ phase spaces of the standard map  along the $\bar{\mathcal{P}}$ direction. Each of the $M$ phase spaces will be referred to as a sub phase space in this paper.

\subsection{Area and effective dimension of eigenstates}
In the CKR model, the Hamiltonian is nonintegrable as long as $K$ is turned nonzero, but even in the region $1<K<5$, there is still a finite portion of quasi-periodic trajectories surviving under the strong kicking strength. Under these $K$, the classical phase space is divided clearly into two kinds of region: some small integrable islands  and  a large chaotic sea \cite{Chirikov2008}. If an initial state lies in the chaotic region, it will explore almost everywhere in the chaotic sea during its long-time dynamics. On the contrary, if an initial state lies in one integrable island, it will remain on one trajectory which forms a 1-dimensional line inside the integrable island. 
Thus we can tell whether a trajectory is integrable or chaotic by its area in the phase space. In practice, we divide the phase space to
$N_c\times N_c$ cells and define the coarse-grained area of a trajectory by the number of cells passed through by the trajectory.
Then the
area of a chaotic trajectory will be proportional to $N_c^2$, while that of an integrable trajectory will be proportional to $N_c$, which gives a rigorous division in the limit $N_c \rightarrow \infty$.

As the quantum phase space is naturally ``coarse-grained" by Planck cells, we can  define the area of an eigenstate,
which serves as a criteria to distinguish integrable and chaotic eigenstates. We define the area $\mathcal{A}$ of a given state $|\psi\rangle$ as
\begin{equation}
\mathcal{A}(|\psi\rangle) = \lr{ \sum_{\mathcal{X},\mathcal{P}} |\langle\mathcal{X},\mathcal{P}|\psi\rangle|^4}^{-1}.
\end{equation}
It is clear that each Wannier basis has area $\mathcal{A}(|\mathcal{X},\mathcal{P}\rangle)=1$; if $|\psi\rangle$ is equally distributed in $N_\psi$ Planck cells while it has no overlap with other cells, its $\mathcal{A}$ will be equal to $N_\psi$.
Thus, this definition can reflect the extent of expansion of the state in the quantum phase space. Note that  this quantity is called the inverse participation ratio defined in a slightly different context \cite{IPR1, IPR2, IPR3, haake, heyl2018quantum}.

We expect in the semiclassical limit $N_x,N_p\rightarrow \infty$ with $N_x/N_p$ constant, $\mathcal{A}\propto N=N_xN_p$ for chaotic eigenstates and $\mathcal{A}\propto \sqrt{N}$ for integrable ones. Since $\hbar_e\propto 1/N$, we define the effective dimension of each eigenstate $\phi$:
\begin{equation}
\label{Deff}
D_{\rm eff}(\phi) = -2\lim_{\hbar_e\rightarrow 0} \frac{\ln \mathcal{A}(\phi)}{\ln \hbar_e}.
\end{equation}
which will be close to $1$ for integrable eigenstates and $2$ for chaotic ones. We note that although $\mathcal{A}$ is dependent on the construction detail of phase space, $D_{\rm eff}$ is universal. Instead of looking at the Husimi distribution of each eigenstate to determine which type that state belongs, we can make the discrimination directly from the value of its area or effective dimension by means of the Wannier phase space, which enables us to make the classification of all eigenstates, just as in classical mechanics where a single Poincar\'e section can depict the behavior of all orbits.

In the definition of $D_{\rm eff}$, one needs to relate eigenstates at different $\hbar_e$. This is not straightforward
as the number of all eigenstates varies with $\hbar_e$. To relate eigenstates,
we sort all eigenstates by their area, and get the index $\ell_{{\cal A}(\phi)}\in\{1,\cdots,N\}$ for each $\phi$. Then we label each $\phi$ by its normalized position $\ell_\phi\equiv \ell_{{\cal A}(\phi)}/N\in[0,1]$. Finally, two states at different $\hbar_e$ are regarded  as the same eigenstate if they have the same $\ell_\phi$.

\begin{figure}[htbp]
\includegraphics[width=\figwid]{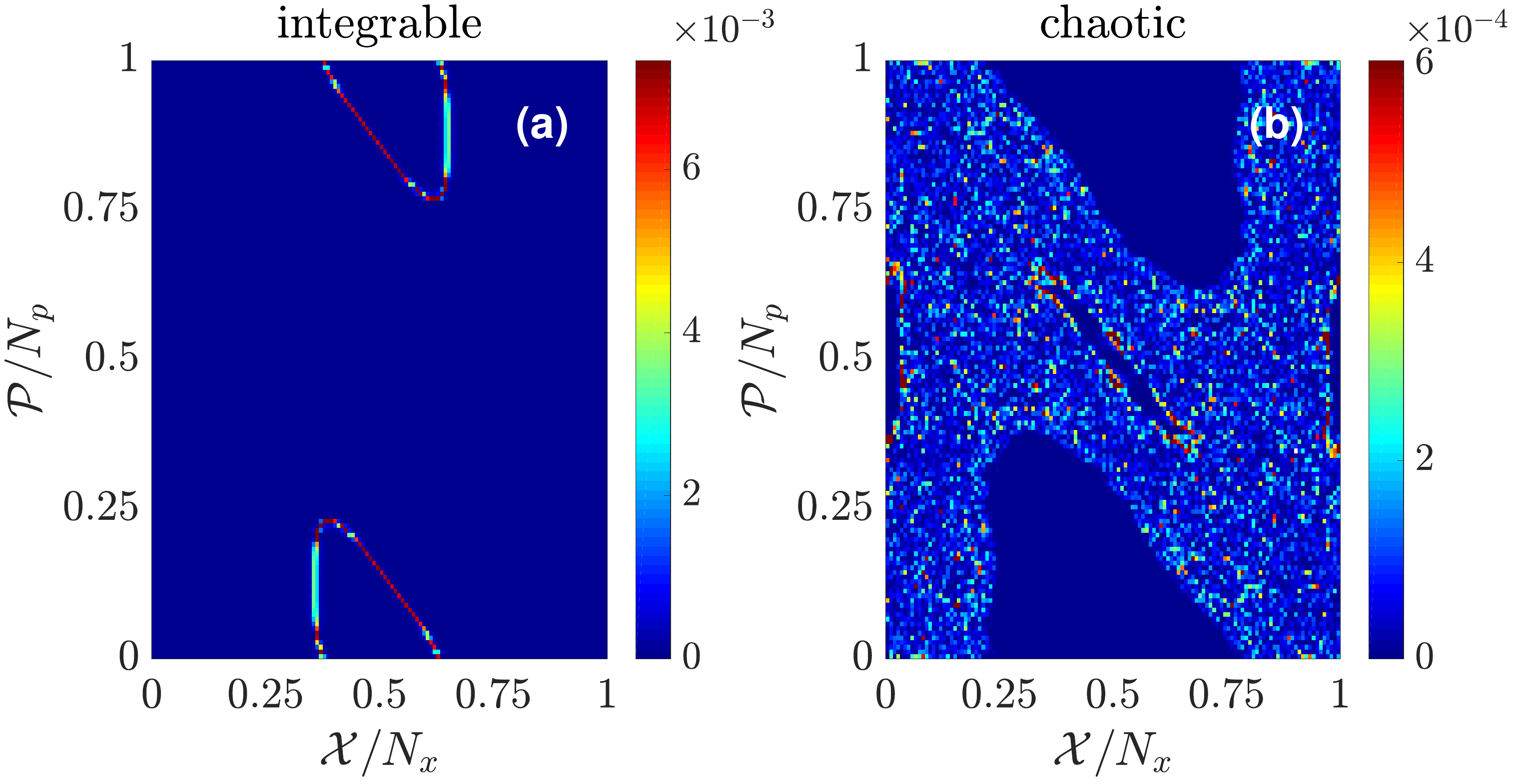}
\caption{\label{fig:projXP} Phase space representation of (a) an integrable eigenstate and (b) a chaotic eigenstate at $K=2, N_x=N_p=128$. The value of each cell is $|\langle\mathcal{X},\mathcal{P}|\phi\rangle|^2$, where $|\phi\rangle$ is the eigenstate.  }
\end{figure}

\section{manifestation of KAM in QKR}

\subsection{Quantum resonance: $\hbar_e=2\pi/N_x^2$ }
In this section, we present our main results using the Wannier basis to investigate the classification of eigenstates in the system. We first consider the simplest case $\hbar_e=2\pi/N_x^2$. As expected, there are two
types of eigenstates, and
two examples are shown in Fig.~\ref{fig:projXP}.

\begin{figure}[!hb]
\includegraphics[width=\figwid]{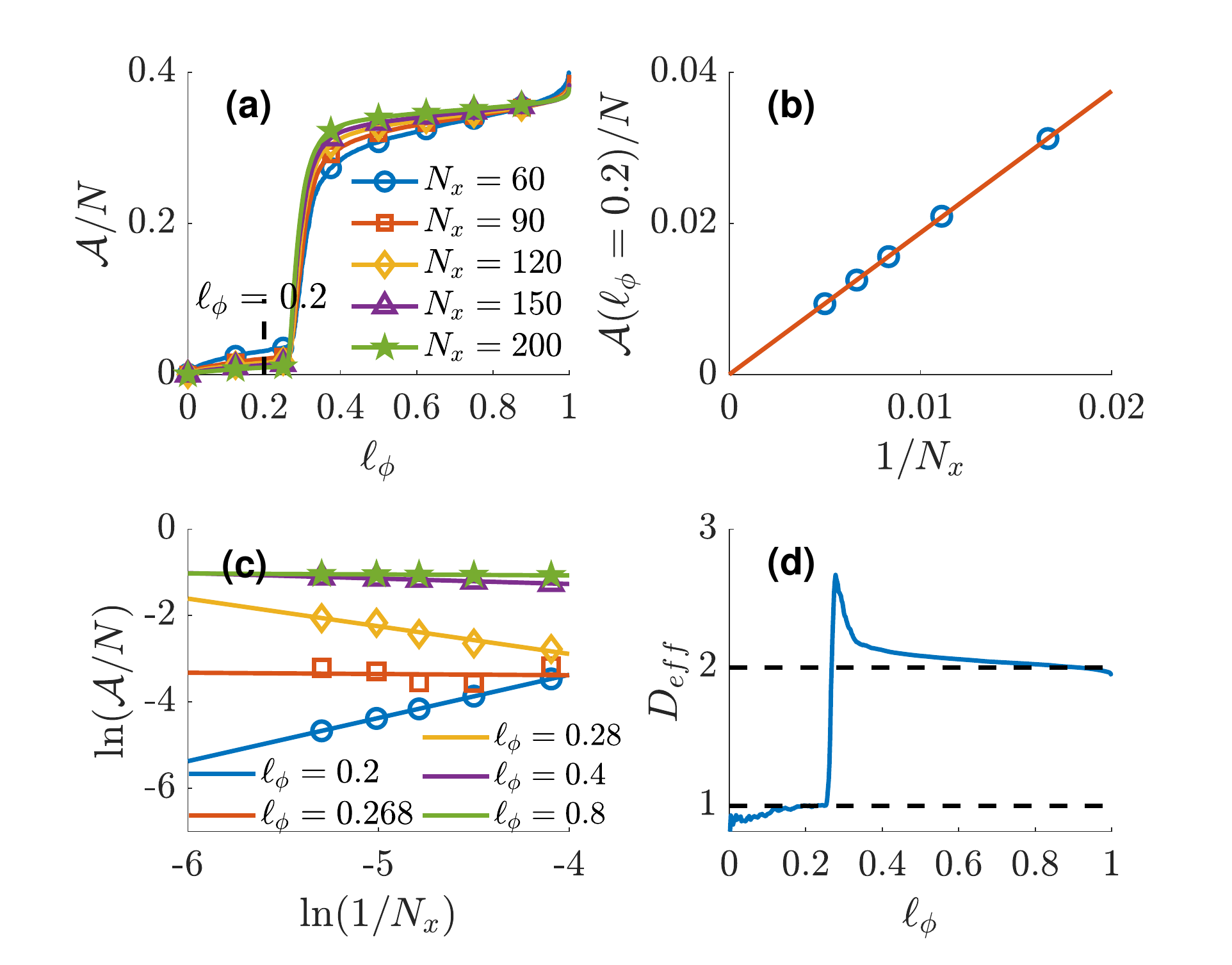}
\caption{\label{fig:areaN} (a) $\mathcal{A}(\phi)$ of all eigenstates at different $N_x$. (b) $\mathcal{A}(\phi)$ at $\ell_\phi=0.2$ for different $N_x$. (c) Logarithmic fitting for 5 typical $\ell_\phi$. (d) The effective dimenstion $D_{\rm eff}$ of all eigenstates $\ell_\phi$,
where $D_{\rm eff}$ is calculated from the slope of the logarithmic fitting. The parameters are $\hbar_e=\frac{2\pi}{N_x^2}, K=2$.}
\end{figure}

We calculate the area $\mathcal{A}$ for each eigenstate, and there is a sharp step when $\mathcal{A}$ is plotted  as a function of
the eigenstate index $\ell_\phi$ (see Fig.~\ref{fig:areaN}(a)). The step gets sharper when
$N_x$ is increased, or equivalently, when $\hbar_e$ is decreased. This sharp step defines
a critical value $\ell_\phi^{c}$. One can roughly say that the eigenstates with $\ell_\phi<\ell_\phi^{c}$ are integrable and those with
$\ell_\phi>\ell_\phi^c$ are chaotic. Moreover, one expects that the area at $\ell_\phi<\ell_\phi^c$ is $\mathcal{A}(\phi)/N\propto 1/N_x$
(see Fig.~\ref{fig:areaN}(b)) while $\mathcal{A}(\phi)/N$ tends to constant at $\ell_\phi >\ell_\phi^c$.

The effective dimension $D_{\rm eff}$ is also calculated and is plotted in Fig.~\ref{fig:areaN}(d).
As expected,  $D_{\rm eff}=1$ for eigenstates below $\ell_\phi^c$ and $D_{\rm eff}=2$ for eigenstates above $\ell_\phi^c$.
However, near $\ell_\phi^c$, $D_{\rm eff}$ deviates from both $1$ and $2$. It may indicate  the existence
of hierarchial states described in Ref.\cite{hierarchy_PhysRevLett.85.1214}. These states correspond to classical orbits which are trapped in the vicinity of the hierarchy of integrable islands for a long time, but will finally leak into the chaotic sea.
These states will disappear when $\hbar_e\rightarrow 0$ \cite{hierarchy_PhysRevLett.85.1214}.

\begin{figure}[!htbp]
\includegraphics[width=\figwid]{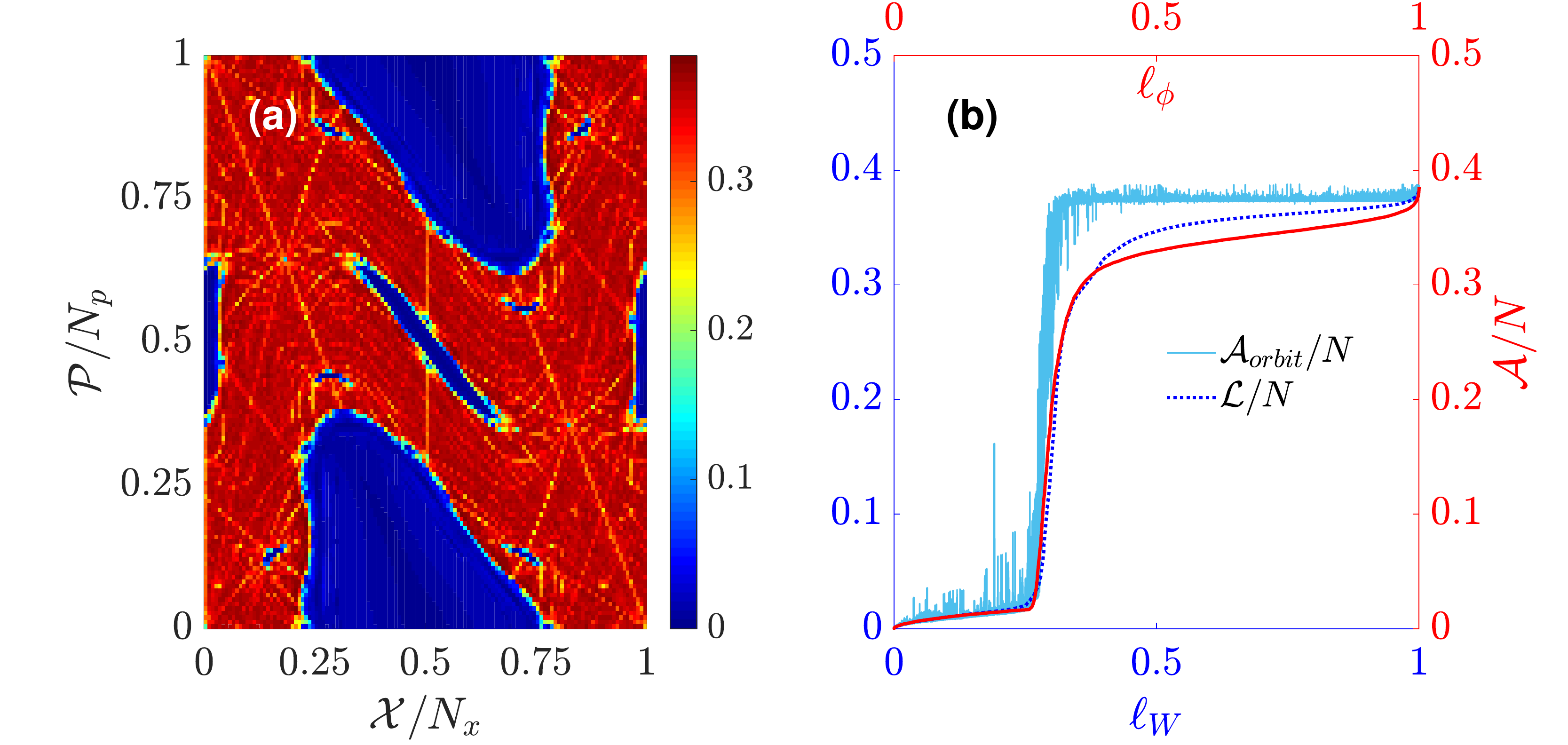}
\caption{\label{fig:XPlen} (a) Length $\mathcal{L}/N$ of each Wannier basis. (b) Sorted area $\mathcal{A}/N$ of each eigenstate (red line), sorted $\mathcal{L}/N$ of each Wannier basis (dark blue dotted line) and long-time area $\mathcal{A}_{orbit}/N$ of them (light blue solid line).  Similar to how eigenstates are sorted and labelled by $\ell_\phi$, each Wannier basis is sorted by its length $\mathcal{L}$ and is then labelled by $\ell_W\in[0,1]$.  The parameters for both sub-figures are  $K=2, N_x=N_p=128$.  }
\end{figure}

We have projected unitarily one set of basis (eigenstates) to another (Wannier basis), which gives information
about how many Planck cells each individual eigenstate occupies. We can reverse the unitary transformation,
and expand  Wannier basis in terms of the eigenstates; the expansion coefficients tell us not only how the Wannier basis form the eigenstates, but 
more importantly how  an initial state localized in the phase space will evolve for a long time.
To illustate this,
we define the length $\mathcal{L}$ of a Planck cell $|\mathcal{X},\mathcal{P}\rangle$ as
\begin{equation}
\mathcal{L}= \left( \sum_{\phi} |\langle\mathcal{X},\mathcal{P}|\phi\rangle|^4\right)^{-1},
\end{equation}
which measures how much $|\mathcal{X},\mathcal{P}\rangle$ occupies in the spectrum. We have computed $\mathcal{L}$ for each Wannier function $|\mathcal{X},\mathcal{P}\rangle$, and
the results are plotted in Fig.~\ref{fig:XPlen}(a), which resembles the classical Poincar\'e section that is divided into integrable and chaotic regions. 
Specifically, it is those Wannier bases in the classical integrable region that have small $\mathcal{L}$, while the others in the classical chaotic region have large $\mathcal{L}$. 

Interestingly, the length $\mathcal{L}$ of  a Planck cell $|\mathcal{X},\mathcal{P}\rangle$  
in fact also indicates how many Planck cells  the system will explore dynamically if it starts at $|\mathcal{X},\mathcal{P}\rangle$. 
To see this, we define the long-time area for a Planck cell $|\mathcal{X},\mathcal{P}\rangle$  as 
\begin{equation}
\mathcal{A}_{orbit}=\lr{\left\langle \sum_{\mathcal{X'},\mathcal{P'}} |\langle\mathcal{X'},\mathcal{P'}|V^{n_T}|\mathcal{X},\mathcal{P} \rangle|^4\right\rangle_{n_T}}^{-1}
\end{equation}
 Here $\langle\cdot\rangle_{n_T}$ means taking the average of $n_T$, the number of periods. In practice we use diagonal ensemble to calculate this value, 
 (see Appendix~\ref{DE} for details). In Fig.~\ref{fig:XPlen}(b), we compare $\mathcal{L}$ of each Wannier function $|\mathcal{X},\mathcal{P}\rangle$ (dark blue dotted line) with its long-time area (light blue solid line). The sorted area of eigenstates (red line) is also plotted. The figure clearly shows that these three curves are close to each other, especially in the integrable part. These results indeed show that the length $\mathcal{L}$ of 
 a given Planck cell $|\mathcal{X},\mathcal{P}\rangle$ measures how much phase space it will explore dynamically. 

\begin{figure}[!htbp]
\includegraphics[width=\figwid]{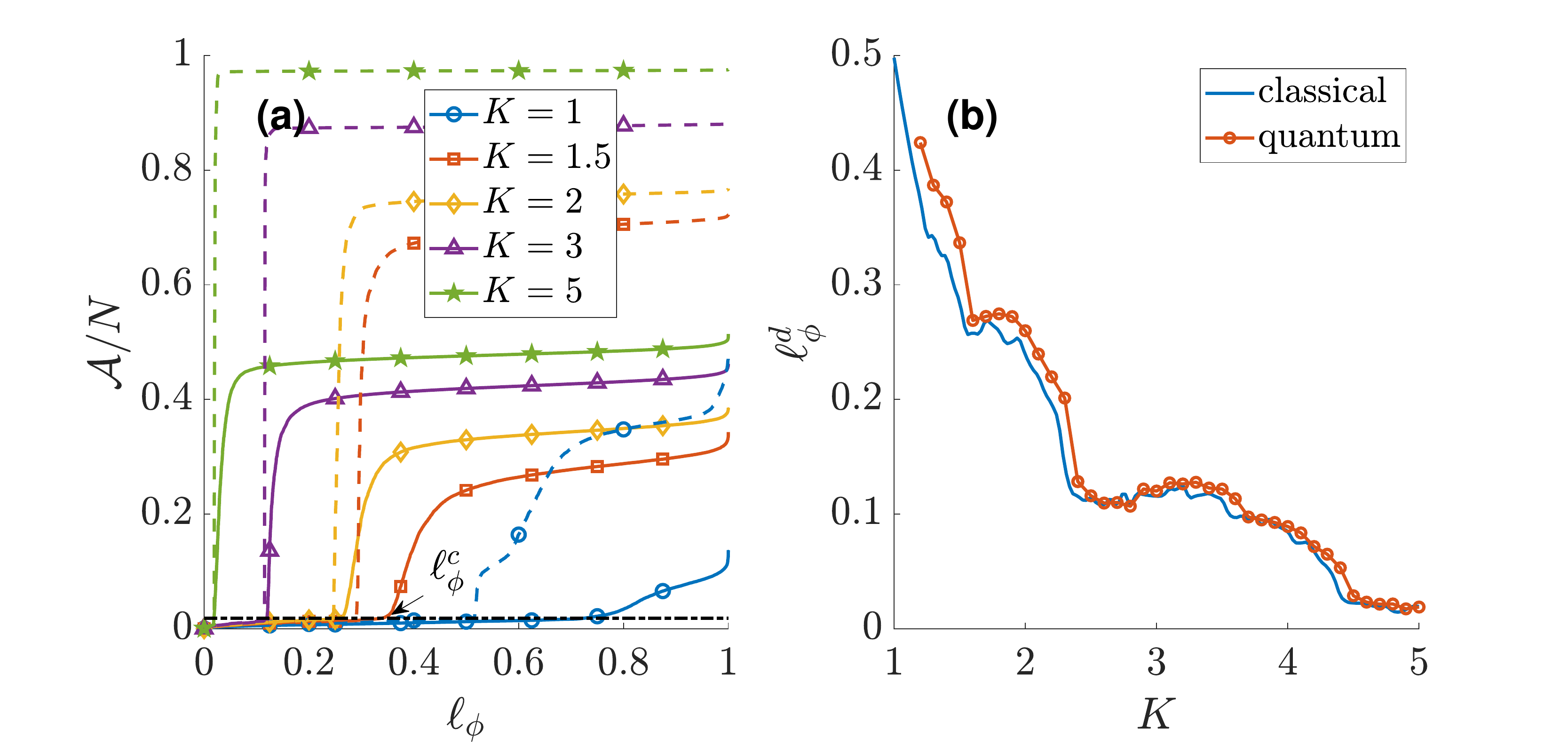}
\caption{\label{fig:areaK} (a) Area of each eigenstate (solid line) and coarse-grained area of
classical trajectories (dashed line).  (b) Demarcation point $\ell_\phi^c$ for classical and quantum cases. $\ell_\phi^c$ is obtained by $\mathcal{A}(\ell_\phi^c)=0.018N, \hbar_e=2\pi/2^{14}$.}
\end{figure}

We now use the Wannier basis to study how KAM breaks down for  increasing $K$. Since the QKR becomes more chaotic as the kicking strength $K$ increases, one expects
that the critical value $\ell_\phi^c$ decreases and eventually becomes zero.
This is indeed the case as shown in  Fig.~\ref{fig:areaK}, where we have also compared these results to their classical counterparts.
For the classial results, we divide the phase space into $N=100\times 100$ cells, choose $10^4$ random initial points and
evolve long enough time ($n_T=10^6$ kicks). Then each trajectory contains $n_T$ points.
For each trajectory, $\mathcal{A}$ is calculated similar to the definition in the quantum case: $\mathcal{A}=\lr{\sum_j (n_j/n_T)^2}^{-1}$, where $n_j$ is the number of points in the $j$th cell. There is great consistency between
the quantum results and the classical results. There are also differences.
 First of all, the saturation value of the classical $\mathcal{A}$ is  much larger and close
 to the area of chaotic sea in the phase space, which indicates that the chaotic sea is classically ergodic.
 The saturation value of the quantum $\mathcal{A}$ is smaller; this is due to the fact that
 the probability distribution of chaotic eigenstates on the phase space has large fluctuations\cite{xiong}.
 Second, the classical demarcation point $\ell_\phi^c$ differs from its quantum counterpart, which means there are more integrable eigenstates in QKR than integrable trajectories in CKR, especially when $K$ is small.
 This is because there are hierarchial states which are supported by the chaotic region but behave like integrable states, as $\hbar_e$ is finite. Moreover, in CKR the hierarchial regions of integrable islands are larger with smaller $K$.

\subsection{Generic $\hbar_e$ and Anderson localization}
In generic cases, $\hbar_e/(2\pi)$ is irrational and the matrix $U$ cannot be reduced to a finite one. However,
we can build a series of rational numbers $M_1/N_1, M_2/N_2,\cdots$,
which has irrational number $\hbar_e/(2\pi)$ as its limit. For each $j$, we have a resonant matrix $U_j$
with effective Planck constant $\hbar_{e,j}=2\pi M_j/N_j$, and we can do the previous reduction
and construct the Wannier phase space. The properties of the system with the original $\hbar_e$
are approximated by increasing $j$.

Without loss of generality, we let $\hbar_e=2\pi / (N_{x} + \Delta)^2$, where $N_{x}$ is
an even integer and $\Delta \in [-1,1)$ is irrational. Then we construct
a series $\{M_j/(N_xN_{p,j})\}$ such that  the series $\{M_j/N_{p,j}\}$ approaches $ N_x/ (N_x + \Delta)^2$.
For this series, the quantum phase space  has $N_j=N_x\cdot N_{p,j}$ Wannier states in total.

\begin{figure}[!htbp]
\includegraphics[width=\figwid]{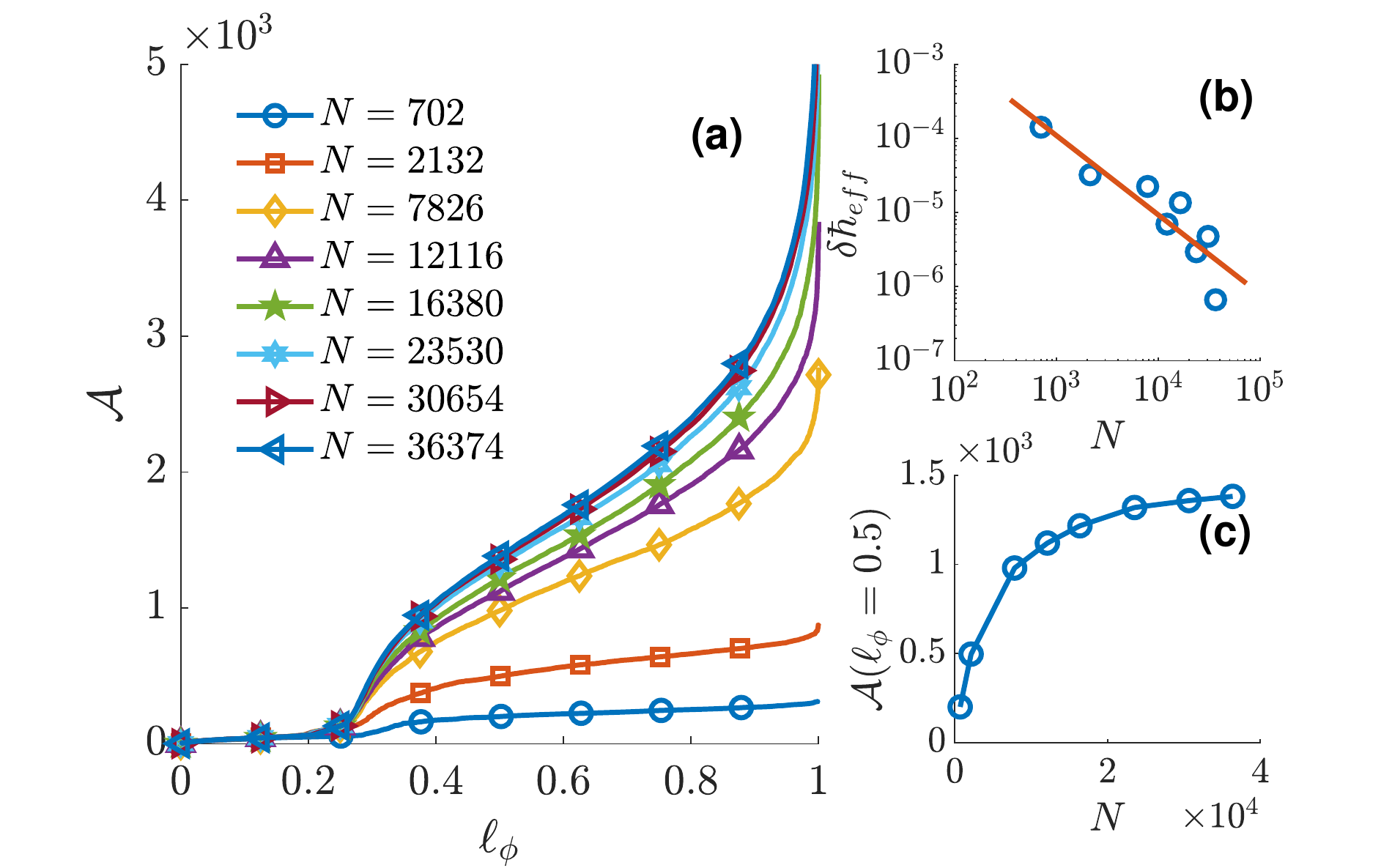}
\caption{\label{fig:areahN} (a) Area $\mathcal{A}$ of eigenstates for each $j$. (b) Rational approximation of generic $\hbar_e$. For each $j$, $\delta \hbar_e\equiv |\hbar_e-\hbar_{e,j}|$. (c) $\mathcal{A}$ at $\phi=0.5$, which saturates when $N\rightarrow \infty$. The parameters are $K=2, N_x = 26, \Delta = 1/\sqrt{2}$, and the index $j$ is omitted.}
\end{figure}

In Fig.~\ref{fig:areahN} we plot the rational approximation of $\hbar_e$ and area of eigenstates for each $j$. The area of integrable eigenstates remains a small constant when $N$ increases, because each eigenstate is confined in one integrable island of one sub phase space, which contains a constant amount of Planck cells. On the other hand, the area of chaotic eigenstates increases with $N$ when $N$ is small, and saturates when $N$ is large enough. The initial growth is consistent with the classical version, in which the chaotic regions of each sub phase space are connected and one point can transport freely in the chaotic sea of the whole phase space. However, the effect of Anderson localization comes in when $N$ is sufficiently large \cite{AL_PhysRevA.29.1639}, which is a pure quantum effect and sets an upper bound of $\mathcal{A}$. To be specific, the localization length in $p$ space of each eigenstate is approximately $n_{loc}=\frac{1}{2}D_c/\hbar_e^2$, where $D_c$ is the classical diffusion coefficient \cite{loc_len_PhysRevLett.56.677}. If $N>n_{loc}$, although the chaotic eigenstates are not confined in one integrable island, they are also localized in some part of the phase space, whose area is of the order of $n_{loc}$ and independent of $N$.

\begin{figure}[!htbp]
\includegraphics[width=\figwid]{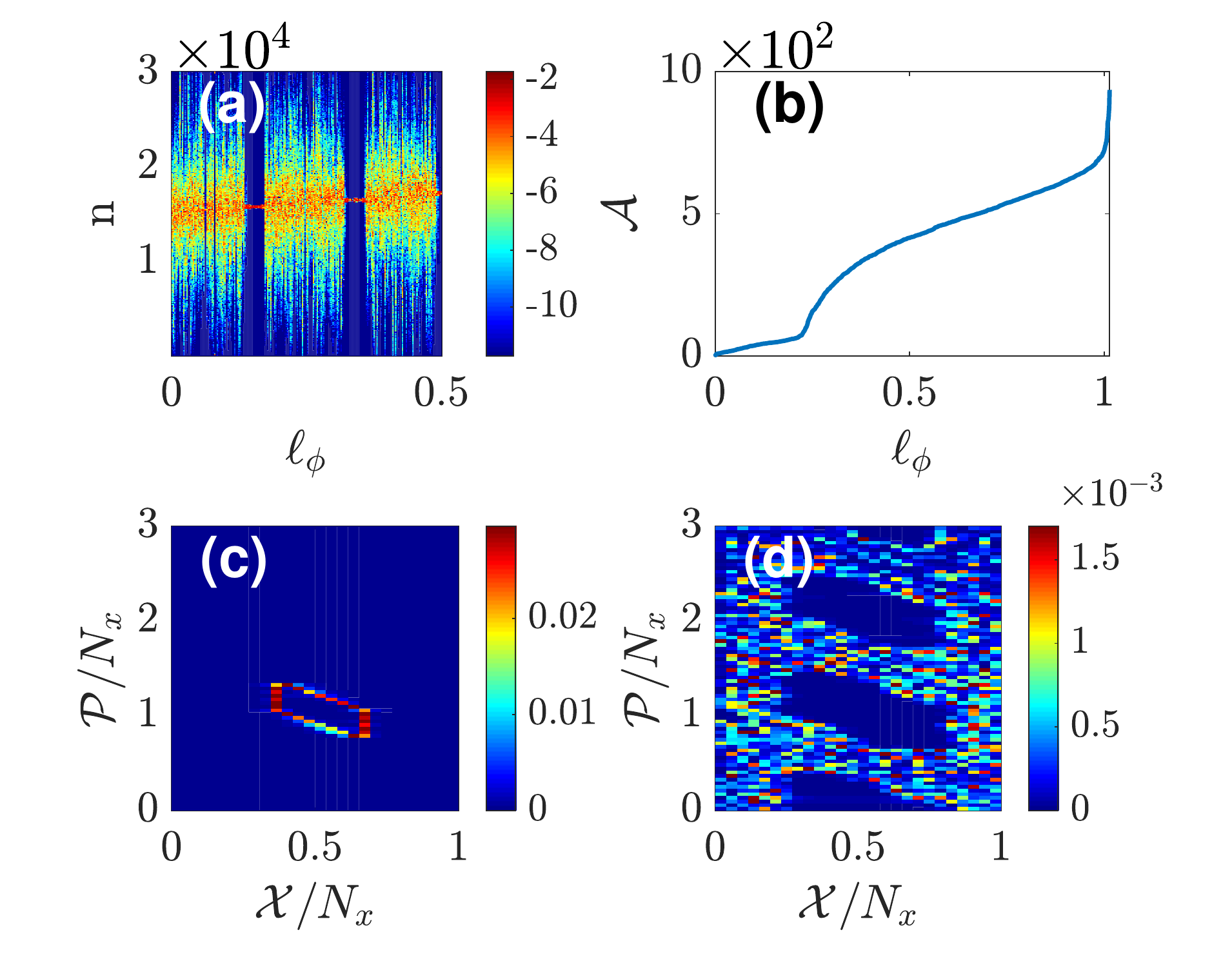}
\caption{\label{fig:irrational} Eigenstates of $U$ with generic $\hbar_e$ without reduction. After diagonalization, we only choose $\sim 2\times 10^3$ eigenstates $\phi$ whose average momentum $\langle n\rangle\sim 1.5\times 10^4$ where the whole $p$ space is $1\le n \le 3\times 10^4$. The quantum phase space is constructed by $3N_x^2$ adjacent momentum eigenstates near $n\sim 1.5\times 10^4$, where $N_x=26$, $\hbar_e=2\pi/(N_x+\Delta)^2, \Delta=1/\sqrt{2},K=2$. (a) $\ln\left|\langle n|\phi\rangle\right|^2$. (b) Area of eigenstates. The value of $\mathcal{A}$ is normalized by the projection of each eigenstate to the phase space: $\mathcal{A}(\phi)= \frac{ \lr{\sum_{\mathcal{X},\mathcal{P}} |\langle\mathcal{X},\mathcal{P}|\phi\rangle|^2 }^2 }{\sum_{\mathcal{X},\mathcal{P}} |\langle\mathcal{X},\mathcal{P}|\phi\rangle|^4} $. (c) One typical integrable eigenstate in quantum phase space. (d) One typical chaotic eigenstate in quantum phase space. }
\end{figure}

For a one step evolution matrix  $U$ with generic $\hbar_e$, we can also simply set a large momentum cutoff $n_{cut}$ ($\gg n_{loc}$) and only consider those eigenstates which are localized in the center of the whole $p$ space. These states have small truncation error, and we are also able to apply the Wannier basis analysis to them. The quantum phase space can be constructed as follows. Choose $N=N_x\cdot N_p$ adjacent momentum eigenstates $|n_0+1\rangle,\cdots,|n_0+N\rangle$, relabel them as $|1\rangle,\cdots,|N\rangle$ and apply Eqs.~(\ref{eqn:XP}) to generate the Wannier basis which constitutes the phase space. The ambiguity here is that the phase space depends on $n_0$, which is insignificant because the change of $n_0$ only causes a slight displacement of Planck cells in the phase space.  In a similar manner, we can project the eigenstates onto the phase space we have constructed. In Fig.~\ref{fig:irrational} we show that these eigenstates are also separated to integrable and chaotic ones, which justifies that this structure of eigenstates depends on neither the previous rational approximation nor the reduction process.

\section{Conclusion}
In sum, we have developed a method based on Wannier phase space to approach KAM
effect in quantum systems. In this approach, each Planck cell in the quantum phase space
is represented by a Wannier function; all the Wannier functions together form a
complete and orthonormal basis.  With the example of QKR, this approach has been shown quite powerful.
First, it has lead us to define the area and effective dimension of eigenstates, which
then give us quantitative measures to divide all eigenstates into integrable and chaotic classes.
Second, it has allowed us define the length of each Planck cell, which measures quantitatively
how many Planck cells the system will traverse if it starts at one Planck cell.
Thirdly, this approach is also used to clarify the distinction between KAM and Anderson localization in QKR. We have used this approach in systems with a classical limit, and it is interesting to ask whether it can be generalized to other quantum systems like spin chains.
This work complements our understanding of the quantum-classical correspondence, and may provide insight into short-wavelength physics such as microcavity photonics.

\begin{acknowledgments}
This work is supported by the The National Key R\&D Program of China (Grants No.~2017YFA0303302, No.~2018YFA0305602), NSFC under Grant No. 11604225 and No. 11734010, Beijing Natural Science Foundation (Z180013), Foundation of Beijing Education Committees under Grant No. KM201710028004.
\end{acknowledgments}

\appendix
\section{\label{reson}Quantum resonance in QKR}
\subsection{Classical origin of the translational invariance}
The existence of quantum resonance in QKR relies on the emergence of a translational invariance in $p$ space, which can be understood in the classical limit\cite{PRAChang}. The classical kicked rotor is described by a pair of the classical conjugate variables $(x_c,p_c)$. Its dynamics is 
an iterating map 
\begin{eqnarray}\label{eq:cla_iter}
p'_c &=& p_c + K_c \sin\lr{2\pi x_c}\,, \\
2\pi x'_c &=& 2\pi x_c + p'_c\,.
\end{eqnarray}
This map  is invariant under the transformation $p_c\rightarrow p_c+2\pi M_c$ where $M_c$ is an integer. In QKR, the angular momentum $p_c$ is quantized, 
that is, $p_c=m\hbar_e$.  Thus the  transformation becomes $m\hbar_e+2\pi M_c=m'\hbar_e$, where  $m,m',M_c$ are integers. It is clear that for QKR being
invariant under this transformation,  $\hbar_e/(2\pi)$ has to be rational.

\subsection{Existence of the Bloch states}
We here use group theory to show the existence of the Bloch states in $p$ space in QKR 
under the condition $U_{n+N\ell, n'+N\ell}=U_{n n'} (l=0,\pm 1,\cdots)$. That is to prove Eqs.~(\ref{eq:bloch},\ref{eq:eigen},\ref{eq:V}).

Let $\mathcal{T}$ be the operator that translate the system in $p$ space by $N$, $\mathcal{T}|n\rangle=|n+N\rangle$. One can prove that 
$\lr{\mathcal{T}U\mathcal{T}^{-1} }_{nn'}=U_{n-N, n'-N}=U_{nn'}$, indicatinng that QKR has a translational symmetry 
$\mathcal{T}$ in $p$ space, similar to  the  translational symmetry in $x$ space for crystal. All operators of the type $\mathcal{T}^k$, where $k$ is an integer, form a symmetry group. Since it is Abelian, each eigenstate $|\phi\rangle$ of $U$ is a one dimensional irreducible representation of the group. That suggests $\mathcal{T}|\phi\rangle=e^{-i\theta}|\phi\rangle$ for some $\theta\in[0,2\pi)$, which leads to Eq.~(\ref{eq:bloch}).

Consider the eigen-equation $\sum_{n'} U_{nn'}\phi(n')=e^{-i\omega_\phi}\phi(n)$. With Eq.~(\ref{eq:bloch}), we have for $s=1,\cdots,N$,
\begin{eqnarray}
e^{-i\omega_\phi}\phi(s) &=& \sum_{n'=-\infty}^\infty U_{sn'}\phi(n') \nonumber\\ &=& \sum_{s'=1}^N \sum_{l'=-\infty}^\infty U_{s(s'+Nl')}\phi(s'+Nl') \nonumber\\ &=& \sum_{s'=1}^N \sum_{l'=-\infty}^\infty U_{s(s'+Nl')}e^{-il'\theta} \phi(s')
\end{eqnarray}
This is just Eqs.~(\ref{eq:eigen},\ref{eq:V}).

\section{Long-time area}\label{DE}

In this section, we provide the details of calculating long-time area of evolved states by diagonal ensemble. Here we consider a general initial state $|\psi_0\rangle=\sum_\phi a_\phi|\phi\rangle$, while previous results are for the special case $|\psi_0\rangle=|\mathcal{X},\mathcal{P}\rangle$. Its inverse area after $n_T$ periods is given by
\begin{eqnarray}
&&\mathcal{A}^{-1}(n_T) = \sum_{\mathcal{X}',\mathcal{P}'} \left| \langle \mathcal{X}',\mathcal{P}'|U^{n_T}|\psi_0\rangle \right|^4 \nonumber\\
&=& \sum_{\mathcal{X}',\mathcal{P}'} \sum_{\phi_1,\phi_2,\phi'_1,\phi'_2} e^{-i n_T(\omega_{\phi_1}+\omega_{\phi_2} - \omega_{\phi'_1} -\omega_{\phi'_2} )} \nonumber\\ && a_{\phi_1}a_{\phi_2} a^*_{\phi'_1} a^*_{\phi'_2}
  \langle \mathcal{X}',\mathcal{P}'|\phi_1\rangle \langle \mathcal{X}',\mathcal{P}'|\phi_2\rangle \nonumber\\ && \langle \mathcal{X}',\mathcal{P}'|\phi'_1\rangle^* \langle \mathcal{X}',\mathcal{P}'|\phi'_2\rangle^*
\end{eqnarray}
Then one can take the average of $n_T$
\begin{eqnarray}
&&\langle e^{-i n_T(\omega_{\phi_1}+\omega_{\phi_2} - \omega_{\phi'_1} -\omega_{\phi'_2} )}\rangle_{n_T} \nonumber \\
&=&\delta_{\phi_1\phi'_1}\delta_{\phi_2\phi'_2}+\delta_{\phi_1\phi'_2}\delta_{\phi_2\phi'_1}
\end{eqnarray}
by assuming that there is no degeneracy in differences of quasi-energies, which is the case in QKR. At last, one gets the diagonal ensemble value
\begin{eqnarray}
\mathcal{A}_{orbit}^{-1} &=& 2\sum_{\mathcal{X'},\mathcal{P'}}\lr{ \sum_{\phi} |a_{\phi}|^2 |\langle \mathcal{X'},\mathcal{P'}|\phi\rangle |^2 }^2 \nonumber\\
&&-\sum_{\mathcal{X'},\mathcal{P'}}\sum_{\phi} |a_\phi|^4 |\langle \mathcal{X'},\mathcal{P'}|\phi\rangle |^4
\end{eqnarray}

\bibliography{QKR_KAM}

\providecommand{\noopsort}[1]{}\providecommand{\singleletter}[1]{#1}%
\begin{thebibliography}{38}
\expandafter\ifx\csname natexlab\endcsname\relax\def\natexlab#1{#1}\fi
\expandafter\ifx\csname bibnamefont\endcsname\relax
  \def\bibnamefont#1{#1}\fi
\expandafter\ifx\csname bibfnamefont\endcsname\relax
  \def\bibfnamefont#1{#1}\fi
\expandafter\ifx\csname citenamefont\endcsname\relax
  \def\citenamefont#1{#1}\fi
\expandafter\ifx\csname url\endcsname\relax
  \def\url#1{\texttt{#1}}\fi
\expandafter\ifx\csname urlprefix\endcsname\relax\def\urlprefix{URL }\fi
\providecommand{\bibinfo}[2]{#2}
\providecommand{\eprint}[2][]{\url{#2}}

\bibitem[{\citenamefont{Arnold}(2013)}]{arnol2013mathematical}
\bibinfo{author}{\bibfnamefont{V.~I.} \bibnamefont{Arnold}},
  \emph{\bibinfo{title}{Mathematical methods of classical mechanics}},
  vol.~\bibinfo{volume}{60} (\bibinfo{publisher}{Springer Science \& Business
  Media}, \bibinfo{year}{2013}).

\bibitem[{\citenamefont{Gutzwiller}(1990)}]{Gutzwiller}
\bibinfo{author}{\bibfnamefont{M.~C.} \bibnamefont{Gutzwiller}},
  \emph{\bibinfo{title}{Chaos in Classical and Quantum Mechanics}},
  vol.~\bibinfo{volume}{60} (\bibinfo{publisher}{Springer, New York},
  \bibinfo{year}{1990}).

\bibitem[{\citenamefont{Kolmogorov}(1954)}]{kolmogorov1954conservation}
\bibinfo{author}{\bibfnamefont{A.}~\bibnamefont{Kolmogorov}}, in
  \emph{\bibinfo{booktitle}{Dokl. Akad. Nauk. SSR}} (\bibinfo{year}{1954}),
  vol.~\bibinfo{volume}{98}, pp. \bibinfo{pages}{2--3}.

\bibitem[{\citenamefont{M\"{o}ser}(1962)}]{Moser}
\bibinfo{author}{\bibfnamefont{J.}~\bibnamefont{M\"{o}ser}},
  \bibinfo{journal}{Nachr. Akad. Wiss. G\"{o}ttingen, II} pp.
  \bibinfo{pages}{1--20} (\bibinfo{year}{1962}).

\bibitem[{\citenamefont{Arnold}(1963)}]{arnold1963vi}
\bibinfo{author}{\bibfnamefont{V.~I.} \bibnamefont{Arnold}},
  \bibinfo{journal}{Russ. Math. Surv.} \textbf{\bibinfo{volume}{18}},
  \bibinfo{pages}{9} (\bibinfo{year}{1963}).

\bibitem[{\citenamefont{Hose and Taylor}(1983)}]{KAM1983PhysRevLett.51.947}
\bibinfo{author}{\bibfnamefont{G.}~\bibnamefont{Hose}} \bibnamefont{and}
  \bibinfo{author}{\bibfnamefont{H.~S.} \bibnamefont{Taylor}},
  \bibinfo{journal}{Phys. Rev. Lett.} \textbf{\bibinfo{volume}{51}},
  \bibinfo{pages}{947} (\bibinfo{year}{1983}),
  \urlprefix\url{https://link.aps.org/doi/10.1103/PhysRevLett.51.947}.

\bibitem[{\citenamefont{Hose et~al.}(1984)\citenamefont{Hose, Taylor, and
  Tip}}]{Hose_1984}
\bibinfo{author}{\bibfnamefont{G.}~\bibnamefont{Hose}},
  \bibinfo{author}{\bibfnamefont{H.~S.} \bibnamefont{Taylor}},
  \bibnamefont{and} \bibinfo{author}{\bibfnamefont{A.}~\bibnamefont{Tip}},
  \bibinfo{journal}{Journal of Physics A: Mathematical and General}
  \textbf{\bibinfo{volume}{17}}, \bibinfo{pages}{1203} (\bibinfo{year}{1984}),
  \urlprefix\url{https://doi.org/10.1088%2F0305-4470%2F17%2F6%2F016}.

\bibitem[{\citenamefont{Geisel et~al.}(1986)\citenamefont{Geisel, Radons, and
  Rubner}}]{KAMbarrier_PhysRevLett.57.2883}
\bibinfo{author}{\bibfnamefont{T.}~\bibnamefont{Geisel}},
  \bibinfo{author}{\bibfnamefont{G.}~\bibnamefont{Radons}}, \bibnamefont{and}
  \bibinfo{author}{\bibfnamefont{J.}~\bibnamefont{Rubner}},
  \bibinfo{journal}{Phys. Rev. Lett.} \textbf{\bibinfo{volume}{57}},
  \bibinfo{pages}{2883} (\bibinfo{year}{1986}),
  \urlprefix\url{https://link.aps.org/doi/10.1103/PhysRevLett.57.2883}.

\bibitem[{\citenamefont{Reichl and Lin}(1987)}]{Reichl1987}
\bibinfo{author}{\bibfnamefont{L.~E.} \bibnamefont{Reichl}} \bibnamefont{and}
  \bibinfo{author}{\bibfnamefont{W.~A.} \bibnamefont{Lin}},
  \bibinfo{journal}{Foundations of Physics} \textbf{\bibinfo{volume}{17}},
  \bibinfo{pages}{689} (\bibinfo{year}{1987}), ISSN \bibinfo{issn}{1572-9516},
  \urlprefix\url{https://doi.org/10.1007/BF01889542}.

\bibitem[{\citenamefont{Lin and Reichl}(1988)}]{PhysRevA.37.3972}
\bibinfo{author}{\bibfnamefont{W.~A.} \bibnamefont{Lin}} \bibnamefont{and}
  \bibinfo{author}{\bibfnamefont{L.~E.} \bibnamefont{Reichl}},
  \bibinfo{journal}{Phys. Rev. A} \textbf{\bibinfo{volume}{37}},
  \bibinfo{pages}{3972} (\bibinfo{year}{1988}),
  \urlprefix\url{https://link.aps.org/doi/10.1103/PhysRevA.37.3972}.

\bibitem[{\citenamefont{Evans}(2004)}]{Evans2004}
\bibinfo{author}{\bibfnamefont{L.~C.} \bibnamefont{Evans}},
  \bibinfo{journal}{Communications in Mathematical Physics}
  \textbf{\bibinfo{volume}{244}}, \bibinfo{pages}{311} (\bibinfo{year}{2004}),
  ISSN \bibinfo{issn}{1432-0916},
  \urlprefix\url{https://doi.org/10.1007/s00220-003-0975-5}.

\bibitem[{\citenamefont{Gr{\'e}bert and Thomann}(2011)}]{Grebert2011}
\bibinfo{author}{\bibfnamefont{B.}~\bibnamefont{Gr{\'e}bert}} \bibnamefont{and}
  \bibinfo{author}{\bibfnamefont{L.}~\bibnamefont{Thomann}},
  \bibinfo{journal}{Communications in Mathematical Physics}
  \textbf{\bibinfo{volume}{307}}, \bibinfo{pages}{383} (\bibinfo{year}{2011}),
  ISSN \bibinfo{issn}{1432-0916},
  \urlprefix\url{https://doi.org/10.1007/s00220-011-1327-5}.

\bibitem[{\citenamefont{Polkovnikov et~al.}(2011)\citenamefont{Polkovnikov,
  Sengupta, Silva, and Vengalattore}}]{Pol_RevModPhys.83.863}
\bibinfo{author}{\bibfnamefont{A.}~\bibnamefont{Polkovnikov}},
  \bibinfo{author}{\bibfnamefont{K.}~\bibnamefont{Sengupta}},
  \bibinfo{author}{\bibfnamefont{A.}~\bibnamefont{Silva}}, \bibnamefont{and}
  \bibinfo{author}{\bibfnamefont{M.}~\bibnamefont{Vengalattore}},
  \bibinfo{journal}{Rev. Mod. Phys.} \textbf{\bibinfo{volume}{83}},
  \bibinfo{pages}{863} (\bibinfo{year}{2011}),
  \urlprefix\url{https://link.aps.org/doi/10.1103/RevModPhys.83.863}.

\bibitem[{\citenamefont{Brandino et~al.}(2015)\citenamefont{Brandino, Caux, and
  Konik}}]{glimmer_PhysRevX.5.041043}
\bibinfo{author}{\bibfnamefont{G.~P.} \bibnamefont{Brandino}},
  \bibinfo{author}{\bibfnamefont{J.-S.} \bibnamefont{Caux}}, \bibnamefont{and}
  \bibinfo{author}{\bibfnamefont{R.~M.} \bibnamefont{Konik}},
  \bibinfo{journal}{Phys. Rev. X} \textbf{\bibinfo{volume}{5}},
  \bibinfo{pages}{041043} (\bibinfo{year}{2015}),
  \urlprefix\url{https://link.aps.org/doi/10.1103/PhysRevX.5.041043}.

\bibitem[{\citenamefont{Percival}(1973)}]{0022-3700-6-9-002}
\bibinfo{author}{\bibfnamefont{I.~C.} \bibnamefont{Percival}},
  \bibinfo{journal}{Journal of Physics B: Atomic and Molecular Physics}
  \textbf{\bibinfo{volume}{6}}, \bibinfo{pages}{L229} (\bibinfo{year}{1973}).

\bibitem[{\citenamefont{Voros}(1976)}]{AIHPA_1976__24_1_31_0}
\bibinfo{author}{\bibfnamefont{A.}~\bibnamefont{Voros}},
  \bibinfo{journal}{Annales de l'I.H.P. Physique th\'eorique}
  \textbf{\bibinfo{volume}{24}}, \bibinfo{pages}{31} (\bibinfo{year}{1976}).

\bibitem[{\citenamefont{Berry}(1977)}]{berry1977}
\bibinfo{author}{\bibfnamefont{M.~V.} \bibnamefont{Berry}},
  \bibinfo{journal}{Journal of Physics A: Mathematical and General}
  \textbf{\bibinfo{volume}{10}}, \bibinfo{pages}{2083} (\bibinfo{year}{1977}).

\bibitem[{\citenamefont{Berry}(1983)}]{berry1983semiclassical}
\bibinfo{author}{\bibfnamefont{M.~V.} \bibnamefont{Berry}},
  \bibinfo{journal}{Les Houches lecture series} \textbf{\bibinfo{volume}{36}},
  \bibinfo{pages}{171} (\bibinfo{year}{1983}).

\bibitem[{\citenamefont{Voros}(1979)}]{voros1979stochastic}
\bibinfo{author}{\bibfnamefont{A.}~\bibnamefont{Voros}}, in
  \emph{\bibinfo{booktitle}{Lecture notes in Physics}}
  (\bibinfo{publisher}{Springer Berlin}, \bibinfo{year}{1979}),
  vol.~\bibinfo{volume}{93}, pp. \bibinfo{pages}{326--333}.

\bibitem[{\citenamefont{Heller}(1984)}]{scar_PhysRevLett.53.1515}
\bibinfo{author}{\bibfnamefont{E.~J.} \bibnamefont{Heller}},
  \bibinfo{journal}{Phys. Rev. Lett.} \textbf{\bibinfo{volume}{53}},
  \bibinfo{pages}{1515} (\bibinfo{year}{1984}).

\bibitem[{\citenamefont{Zyczkowski}(1987)}]{Husimi_PhysRevA.35.3546}
\bibinfo{author}{\bibfnamefont{K.}~\bibnamefont{Zyczkowski}},
  \bibinfo{journal}{Phys. Rev. A} \textbf{\bibinfo{volume}{35}},
  \bibinfo{pages}{3546} (\bibinfo{year}{1987}).

\bibitem[{\citenamefont{Ketzmerick
  et~al.}(2000{\natexlab{a}})\citenamefont{Ketzmerick, Hufnagel, Steinbach, and
  Weiss}}]{newclass_PhysRevLett.85.1214}
\bibinfo{author}{\bibfnamefont{R.}~\bibnamefont{Ketzmerick}},
  \bibinfo{author}{\bibfnamefont{L.}~\bibnamefont{Hufnagel}},
  \bibinfo{author}{\bibfnamefont{F.}~\bibnamefont{Steinbach}},
  \bibnamefont{and} \bibinfo{author}{\bibfnamefont{M.}~\bibnamefont{Weiss}},
  \bibinfo{journal}{Phys. Rev. Lett.} \textbf{\bibinfo{volume}{85}},
  \bibinfo{pages}{1214} (\bibinfo{year}{2000}{\natexlab{a}}).

\bibitem[{\citenamefont{Han and Wu}(2015)}]{HanPRE}
\bibinfo{author}{\bibfnamefont{X.}~\bibnamefont{Han}} \bibnamefont{and}
  \bibinfo{author}{\bibfnamefont{B.}~\bibnamefont{Wu}}, \bibinfo{journal}{Phys.
  Rev. E} \textbf{\bibinfo{volume}{91}}, \bibinfo{pages}{062106}
  (\bibinfo{year}{2015}).

\bibitem[{\citenamefont{Fang et~al.}(2018)\citenamefont{Fang, Wu, and
  Wu}}]{Wannier1742-5468-2018-2-023113}
\bibinfo{author}{\bibfnamefont{Y.}~\bibnamefont{Fang}},
  \bibinfo{author}{\bibfnamefont{F.}~\bibnamefont{Wu}}, \bibnamefont{and}
  \bibinfo{author}{\bibfnamefont{B.}~\bibnamefont{Wu}},
  \bibinfo{journal}{Journal of Statistical Mechanics: Theory and Experiment}
  \textbf{\bibinfo{volume}{2018}}, \bibinfo{pages}{023113}
  (\bibinfo{year}{2018}).

\bibitem[{\citenamefont{{Jiang} et~al.}(2017)\citenamefont{{Jiang}, {Chen}, and
  {Wu}}}]{Jiang}
\bibinfo{author}{\bibfnamefont{J.}~\bibnamefont{{Jiang}}},
  \bibinfo{author}{\bibfnamefont{Y.}~\bibnamefont{{Chen}}}, \bibnamefont{and}
  \bibinfo{author}{\bibfnamefont{B.}~\bibnamefont{{Wu}}},
  \bibinfo{journal}{ArXiv e-prints}  (\bibinfo{year}{2017}),
  \eprint{1712.04533}.

\bibitem[{\citenamefont{Chirikov}(1979)}]{CHIRIKOV1979263}
\bibinfo{author}{\bibfnamefont{B.~V.} \bibnamefont{Chirikov}},
  \bibinfo{journal}{Physics Reports} \textbf{\bibinfo{volume}{52}},
  \bibinfo{pages}{263 } (\bibinfo{year}{1979}).

\bibitem[{\citenamefont{Chang and Shi}(1986)}]{PRAChang}
\bibinfo{author}{\bibfnamefont{S.-J.} \bibnamefont{Chang}} \bibnamefont{and}
  \bibinfo{author}{\bibfnamefont{K.-J.} \bibnamefont{Shi}},
  \bibinfo{journal}{Phys. Rev. A} \textbf{\bibinfo{volume}{34}},
  \bibinfo{pages}{7} (\bibinfo{year}{1986}).

\bibitem[{\citenamefont{Casati et~al.}(1979)\citenamefont{Casati, Chirikov,
  Izraelev, and Ford}}]{qr_casati}
\bibinfo{author}{\bibfnamefont{G.}~\bibnamefont{Casati}},
  \bibinfo{author}{\bibfnamefont{B.~V.} \bibnamefont{Chirikov}},
  \bibinfo{author}{\bibfnamefont{F.~M.} \bibnamefont{Izraelev}},
  \bibnamefont{and} \bibinfo{author}{\bibfnamefont{J.}~\bibnamefont{Ford}}, in
  \emph{\bibinfo{booktitle}{Stochastic Behavior in Classical and Quantum
  Hamiltonian Systems}}, edited by
  \bibinfo{editor}{\bibfnamefont{G.}~\bibnamefont{Casati}} \bibnamefont{and}
  \bibinfo{editor}{\bibfnamefont{J.}~\bibnamefont{Ford}}
  (\bibinfo{publisher}{Springer Berlin Heidelberg}, \bibinfo{address}{Berlin,
  Heidelberg}, \bibinfo{year}{1979}), pp. \bibinfo{pages}{334--352}.

\bibitem[{\citenamefont{Chirikov and Shepelyansky}(2008)}]{Chirikov2008}
\bibinfo{author}{\bibfnamefont{B.}~\bibnamefont{Chirikov}} \bibnamefont{and}
  \bibinfo{author}{\bibfnamefont{D.}~\bibnamefont{Shepelyansky}},
  \bibinfo{journal}{Scholarpedia} \textbf{\bibinfo{volume}{3}},
  \bibinfo{pages}{3550} (\bibinfo{year}{2008}), \bibinfo{note}{revision
  \#186605}.

\bibitem[{\citenamefont{Jacquod and Shepelyansky}(1995)}]{IPR1}
\bibinfo{author}{\bibfnamefont{P.}~\bibnamefont{Jacquod}} \bibnamefont{and}
  \bibinfo{author}{\bibfnamefont{D.~L.} \bibnamefont{Shepelyansky}},
  \bibinfo{journal}{Phys. Rev. Lett.} \textbf{\bibinfo{volume}{75}},
  \bibinfo{pages}{3501} (\bibinfo{year}{1995}),
  \urlprefix\url{https://link.aps.org/doi/10.1103/PhysRevLett.75.3501}.

\bibitem[{\citenamefont{Fyodorov and Mirlin}(1995)}]{IPR2}
\bibinfo{author}{\bibfnamefont{Y.~V.} \bibnamefont{Fyodorov}} \bibnamefont{and}
  \bibinfo{author}{\bibfnamefont{A.~D.} \bibnamefont{Mirlin}},
  \bibinfo{journal}{Phys. Rev. B} \textbf{\bibinfo{volume}{52}},
  \bibinfo{pages}{R11580} (\bibinfo{year}{1995}),
  \urlprefix\url{https://link.aps.org/doi/10.1103/PhysRevB.52.R11580}.

\bibitem[{\citenamefont{Georgeot and Shepelyansky}(1997)}]{IPR3}
\bibinfo{author}{\bibfnamefont{B.}~\bibnamefont{Georgeot}} \bibnamefont{and}
  \bibinfo{author}{\bibfnamefont{D.~L.} \bibnamefont{Shepelyansky}},
  \bibinfo{journal}{Phys. Rev. Lett.} \textbf{\bibinfo{volume}{79}},
  \bibinfo{pages}{4365} (\bibinfo{year}{1997}),
  \urlprefix\url{https://link.aps.org/doi/10.1103/PhysRevLett.79.4365}.

\bibitem[{\citenamefont{Haake and Haken}(2010)}]{haake}
\bibinfo{author}{\bibfnamefont{F.}~\bibnamefont{Haake}} \bibnamefont{and}
  \bibinfo{author}{\bibfnamefont{H.}~\bibnamefont{Haken}},
  \emph{\bibinfo{title}{Quantum signatures of chaos}}, vol.~\bibinfo{volume}{2}
  (\bibinfo{publisher}{Springer}, \bibinfo{year}{2010}).

\bibitem[{\citenamefont{Heyl et~al.}(2018)\citenamefont{Heyl, Hauke, and
  Zoller}}]{heyl2018quantum}
\bibinfo{author}{\bibfnamefont{M.}~\bibnamefont{Heyl}},
  \bibinfo{author}{\bibfnamefont{P.}~\bibnamefont{Hauke}}, \bibnamefont{and}
  \bibinfo{author}{\bibfnamefont{P.}~\bibnamefont{Zoller}},
  \bibinfo{journal}{arXiv preprint arXiv:1806.11123}  (\bibinfo{year}{2018}).

\bibitem[{\citenamefont{Ketzmerick
  et~al.}(2000{\natexlab{b}})\citenamefont{Ketzmerick, Hufnagel, Steinbach, and
  Weiss}}]{hierarchy_PhysRevLett.85.1214}
\bibinfo{author}{\bibfnamefont{R.}~\bibnamefont{Ketzmerick}},
  \bibinfo{author}{\bibfnamefont{L.}~\bibnamefont{Hufnagel}},
  \bibinfo{author}{\bibfnamefont{F.}~\bibnamefont{Steinbach}},
  \bibnamefont{and} \bibinfo{author}{\bibfnamefont{M.}~\bibnamefont{Weiss}},
  \bibinfo{journal}{Phys. Rev. Lett.} \textbf{\bibinfo{volume}{85}},
  \bibinfo{pages}{1214} (\bibinfo{year}{2000}{\natexlab{b}}).

\bibitem[{\citenamefont{Xiong and Wu}(2011)}]{xiong}
\bibinfo{author}{\bibfnamefont{H.~W.} \bibnamefont{Xiong}} \bibnamefont{and}
  \bibinfo{author}{\bibfnamefont{B.}~\bibnamefont{Wu}}, \bibinfo{journal}{Laser
  Physics Letters} \textbf{\bibinfo{volume}{8}}, \bibinfo{pages}{398}
  (\bibinfo{year}{2011}).

\bibitem[{\citenamefont{Grempel et~al.}(1984)\citenamefont{Grempel, Prange, and
  Fishman}}]{AL_PhysRevA.29.1639}
\bibinfo{author}{\bibfnamefont{D.~R.} \bibnamefont{Grempel}},
  \bibinfo{author}{\bibfnamefont{R.~E.} \bibnamefont{Prange}},
  \bibnamefont{and} \bibinfo{author}{\bibfnamefont{S.}~\bibnamefont{Fishman}},
  \bibinfo{journal}{Phys. Rev. A} \textbf{\bibinfo{volume}{29}},
  \bibinfo{pages}{1639} (\bibinfo{year}{1984}).

\bibitem[{\citenamefont{Shepelyansky}(1986)}]{loc_len_PhysRevLett.56.677}
\bibinfo{author}{\bibfnamefont{D.~L.} \bibnamefont{Shepelyansky}},
  \bibinfo{journal}{Phys. Rev. Lett.} \textbf{\bibinfo{volume}{56}},
  \bibinfo{pages}{677} (\bibinfo{year}{1986}).

\end{thebibliography}

\end{document}